\documentclass[a4paper]{article}
\usepackage{INTERSPEECH2018}
\usepackage[hyphens,spaces,obeyspaces]{url}
\usepackage{xcolor}
\usepackage{soul}
\usepackage{textgreek}
\usepackage[bottom]{footmisc}
\usepackage{graphicx}
\usepackage{amssymb,amsmath,bm}
\usepackage{textcomp}
\usepackage{graphicx}
\usepackage{wrapfig}
\usepackage{lipsum}
\usepackage{dblfloatfix}
\usepackage{fixltx2e}
\usepackage{nameref}
\usepackage{hyperref}
\usepackage{graphicx}
\usepackage{algorithm}
\usepackage{varioref}
\usepackage{booktabs}  
\usepackage{siunitx}
\usepackage[noend]{algpseudocode}
\usepackage{numprint}
\usepackage{textcomp}
\usepackage[T1]{fontenc} 
\usepackage{algorithm}
\usepackage{varioref}
\usepackage{siunitx}
\usepackage{numprint}
\usepackage{textcomp}
\def\BState{\State\hskip-\ALG@thistlm}
\def\vec#1{\ensuremath{\bm{{#1}}}}

\usepackage{lipsum}
\usepackage[noend]{algpseudocode}
\usepackage{tablefootnote}
\makeatletter
\def\BState{\State\hskip-\ALG@thistlm}
\makeatother
\usepackage{xpatch}
\xpatchcmd{\algorithmic}
{\ALG@tlm\z@}{\leftmargin\z@\ALG@tlm\z@}
{}{}
\title{Robust Speaker Clustering using Mixtures of von Mises-Fisher Distributions for Naturalistic Audio Streams}
\name{Harishchandra Dubey, Abhijeet Sangwan, John H. L. Hansen~\thanks{\textcolor{blue}{This material is presented to ensure timely dissemination of scholarly and technical work. Copyright and all rights therein are retained by the authors or by the respective copyright holders. The original citation of this paper is: H. Dubey,  A. Sangwan, J. H. L. Hansen, "Robust Speaker Clustering using Mixtures of von Mises-Fisher Distributions for Naturalistic Audio Streams", ISCA INTERSPEECH, Sept. 2-6, 2018, Hyderabad, India.}}}
\address{
	Robust Speech Technologies Lab, Center for Robust Speech Systems\\
	The University of Texas at Dallas, 
	Richardson, TX- 75080, USA
}
\email{\{Harishchandra.Dubey, Abhijeet.Sangwan, John.Hansen\}@utdallas.edu}
\begin{document}
	\renewcommand{\normalsize}{\fontsize{9}{11}\selectfont}
	\normalsize
	\maketitle
	%
	\begin{abstract}
		%
		Speaker Diarization (i.e. determining who spoke and when?) for multi-speaker naturalistic interactions such as Peer-Led Team Learning (PLTL) sessions is a challenging task. In this study, we propose robust speaker clustering based on mixture of multivariate von Mises-Fisher distributions. Our diarization pipeline has two stages: (i) ground-truth segmentation; (ii) proposed speaker clustering. The ground-truth speech activity information is used for extracting i-Vectors from each speech-segment. We post-process the i-Vectors with principal component analysis for dimension reduction followed by length-normalization. Normalized i-Vectors are high-dimensional unit vectors possessing discriminative directional characteristics. We model the normalized i-Vectors with a  mixture model consisting of multivariate von Mises-Fisher distributions. K-means clustering with cosine distance is chosen as baseline approach. The evaluation data is derived from: (i) CRSS-PLTL corpus; and (ii) three-meetings subset of AMI corpus. The CRSS-PLTL data contain audio recordings of PLTL sessions which is student-led STEM education paradigm. Proposed approach is consistently better than baseline leading to upto 44.48\% and 53.68\% relative improvements for PLTL and AMI corpus, respectively.
	\end{abstract}
	\section{Introduction}
	\label{sec:intro}
	%
	Speaker Diarization attempts to answer~\textit{who spoke and when?} in an audio stream~\cite{tranter2006overview}. Domains involving practical application of speaker diarization are understanding and transcription of broadcast news, audio-recorded meetings, telephonic conversations~\emph{etc}~\cite{huijbregts2012large,hansen2018utdallas}. The challenges in speaker diarization is application-dependent. NIST Rich Transcription evaluations focused on broadcast news and meetings audio while NIST SRE evaluations had summed-channel telephone data~\cite{anguera2012speaker}. Speaker diarization pipeline consists of several components such as speech activity detection (SAD)~\cite{dubey2018icassp,dubey2018tasl}, speaker change detection, speaker clustering, and re-segmentation~\cite{anguera2012speaker}. Among these, speaker clustering is the main component. 
	
	This paper propose a model-based speaker clustering using i-Vectors. The target application is naturalistic audio streams from Peer-Led Team Learning (PLTL) sessions~\cite{dubey2017utdallas,dubey2016interspeech}. PLTL is a student-led STEM education model where a peer-leader facilitate problem-solving among 6-8 students~\cite{dubey2018pltlis,dubey2016slt}. Previously, we developed speech systems for automatic extraction of behavioral metrics related to PLTL sessions such as dominance, word-count and student participation~\emph{etc}~\cite{dubey2016slt,dubey2017csl}. 
	\section{Background}
	Given the importance of robust clustering in speaker diarization, several approaches were developed such as agglomerative hierarchical clustering (AHC)~\cite{sun2010speaker}, top-down clustering~\cite{meignier2006step}, Spherical K-means clustering~\cite{hansen2016prof}~\emph{etc}. Authors proposed joint speaker segmentation and clustering scheme~\cite{anguera2012speaker}. In~\cite{zhu2005combining}, the MAP-adapted Gaussian mixture-models (GMMs) were combined with Bayesian information criterion (BIC) for speaker diarization. A reduced complexity clustering approach leverages modified integer linear programming (ILP)~\cite{dupuy2014recent}. Unsupervised calibration of PLDA scores was used for i-Vector clustering on CALLHOME corpus~\cite{sell2014speaker}. Previously, von Mises-Fisher distribution were used for text-independent speaker identification based on line spectral frequencies (LSFs) features~\cite{taghia2013mises}. The von Mises-Fisher distributions were employed in several applications such as similarity measure for text-snippets~\cite{sahami2006web}, bio-informatics~\cite{mardia2007protein}~\emph{etc}.
	%
	%
	\section{Front-end}
	In this study, our diarization pipeline consists of three stages: (i) speech dereverberation; (ii) ground-truth speech segmentation; and (iii) proposed speaker clustering based on mixture of von Mises-Fisher distributions (movMF). As the focus on this paper is to develop robust speaker clustering, we used ground-truth speech segmentation information. Using ground-truth speech activity detection is important to prevent irrelevant errors due to incorrect segmentation. Previously, researchers found that speaker clustering could be developed independent of other components in diarization pipeline~\cite{sinclair2013challenges}. 
	%
	\subsection{Speech Enhancement and Ground-truth Segmentation}
	%
	The CRSS-PLTL data is significantly reverberated (see section~\ref{sec:pltl_data}) so we perform experiments with both original (raw) and de-reverb audio. We employed weighted prediction error (WPE)-based dereverberation approach~\cite{yoshioka2011blind}. After dereverberation, we get the speaker segments using the ground-truth speech activity labels. From each segment, i-Vector was extracted followed by PCA dimension reduction and length-normalization. We used only raw audio from AMI meeting corpus to avoid reporting too many results. 
	%
	%
	%
	\subsection{i-Vector Speaker Model}
	\label{sec:ivector}
	%
	%
	%
	Diarization involve extracting i-Vectors from short speech segments (typically one second) unlike speaker verification where complete utterance is used. Numerous techniques were developed for clustering i-Vectors using cosine similarity~\cite{senoussaoui2014study,castaldo2008stream}. The i-Vector framework combined the speaker and channel variability sub-spaces of linear distortion model into a total-variability space represented by matrix $\mathbf{T}$~\cite{dehak2011front,hansen2015speaker}. A speaker-and-session-dependent GMM super-vector, $\mathbf{S}$ is decomposed as
	%
	\begin{equation}
		\mathbf{S} = \mathbf{S_{ubm}} + \mathbf{T} \mathbf{w}, 
	\end{equation}
	where $\mathbf{S_{ubm}}$ is the Universal Background Model (UBM) super-vector~\cite{dehak2011front}. The latent variables, $\mathbf{w}$ are i-Vectors. The total-variability matrix $\mathbf{T}$ is a low-rank projection matrix that maps high-dimensional speaker super-vectors to low-dimensional total-variability space~\cite{dehak2011front,hansen2015speaker}. We use frame-level 20-MFCC features extracted from 40ms windows at 10ms skip-rate. A UBM with 512 components was trained for i-Vector extraction~\cite{dehak2011front}. Given the short speaker-segments in PLTL, we choose the i-Vector dimension as 75. We post-processed the segment-level i-Vectors with PCA for dimension reduction followed by length-normalization~\cite{garcia2011analysis}. For rest of this paper, normalized i-Vectors refer to length-normalized ones. 
	\setlength{\textfloatsep}{10pt}
	\noindent
	\begin{algorithm}[!t]
		\caption{\strut \text{Proposed Speaker Clustering}}
		\textbf{Input:} (1) Set $\bm{\chi}$ of $n$ segment-level normalized i-Vectors of dimensions $d$ from complete audio recording; (2) Number of clusters, $N_{c}$.\\
		\textbf{Output:} (1) A disjoint partitioning of $\bm{\chi}$ into $N_{c}$ clusters; (2) Model parameters of mixture of $N_{c}$ $d$-variate vMF distributions. 
		\begin{algorithmic}[1]	
			\Statex \hspace{-6mm} \textbf{METHOD:}		
			\State \textbf{} \textbf{} Initialize all $\alpha_{h}$, $\bm{\mu_{h}}$, $\kappa_{h}$ for $h = 1,\cdots , N_{c}$
			\State \textbf{} \textbf{Repeat} 
			\State \textbf{ }  \{The hardened Expectation-step of EM\}
			\State \textbf{ }  \textbf{ for} $i$ = 1 to $n$ \textbf{do}
			\State \hspace{6 mm} \textbf{  for} $h$ = 1 to $N_{c}$ \textbf{do}
			\State \hspace{9 mm} $f_{h}(\bm{\chi_{i}} | \bm{\theta_{h}}) \leftarrow c_{d}(\kappa_{h})e^{\kappa_{h} \bm{\mu_{h}}^{T} \bm{\chi_{i}}} $
			\State \hspace{6 mm} \textbf{  end-for} 
			\State \hspace{6 mm} \textbf{  for} $h$ = 1 to $N_{c}$ \textbf{do}
			\State \hspace{2 mm} The hardened-distribution of hidden-variables is given by $ q(h | \bm{\chi_{i}}, \bm{\Theta}) \leftarrow  
			\begin{cases}
			1, & \text{if } h = \underset{h'}{\arg\max} $ $ \alpha_{h'} f_{h'}(\bm{\chi_{i}} | \bm{\theta_{h'}}) \\
			0,              & \text{otherwise}
			\label{hard_movmf_update}
			\end{cases}	
			$
			\State \hspace{6 mm} \textbf{   end-for} 
			\State \textbf{ }  \textbf{ end-for} 
			\State \textbf{ }  \{The Maximization-step of EM\}
			\State \hspace{6 mm} \textbf{ for} $h$ = 1 to $N_{c}$ \textbf{do}
			\State \hspace{9 mm} $\alpha_{h} \leftarrow \frac{1}{n} \sum_{i=1}^{n} q(h | \bm{\chi_{i}}, \bm{\Theta)}$
			\State \hspace{9 mm} $\bm{\mu_{h}} \leftarrow \sum_{i=1}^{n} \bm{\chi_{i} } \text{ } q(h|\bm{\chi_{i}}, \bm{\Theta)}$
			\State \hspace{9 mm} $\bar{r} \leftarrow \frac{||\bm{\mu_{h}}||}{n \alpha_{h}}  $
			\State \hspace{9 mm} $ \bm{\mu_{h}} \leftarrow \frac{\bm{\mu_{h}}}{||\bm{\mu_{h}}||}  $
			\State \hspace{9 mm} $ \kappa_{h} \leftarrow \frac{\bar{r}d - \bar{r}^{3}}{1 - \bar{r}^{2}} $
			\State \hspace{6 mm} \textbf{ end-for} 
			\State \textbf{Until} \textit{convergence}
		\end{algorithmic}
		\label{hard_movmf_algo}
	\end{algorithm}	
	\section{Proposed Speaker Clustering}
	Speaker clustering for PLTL sessions is a challenging task due to following reasons: (i) overlapped-speech; (ii) skewed clusters in feature space; (iii) significant reverberation and multiple noise sources~\emph{etc}. We input the number of clusters in proposed approach. We model normalized i-Vectors from an audio stream with a mixture of $N_{c}$ multivariate ($d$-variate) von Mises-Fisher distributions (movMF). Here, $N_{c}$ is number of speakers and $d$ is the i-Vector dimension. Previously, researchers analyzed the advantages of i-Vector normalization for speaker modelling~\cite{garcia2011analysis}. Normalized i-vectors are high dimensional data lying on unit hypersphere. There are no closed form solutions for movMF parameters. However, it is possible to get reliable estimates of movMF parameters if input data is high-dimensional~\cite{banerjee2005clustering}. Normalized i-Vectors have significantly higher dimensions (e.g. 75) as compared to number of speaker (e.g. 8), thus movMF model could be approximated reliably for i-Vector clustering. The vMF distribution defines a probability density function (PDF) of data lying on unit hypersphere. For modeling the normalized i-Vectors with movMF model, we have a weight parameter, $\alpha$ for each vMF distribution. We adopt expectation-maximization (EM) algorithm for approximating the maximum likelihood (ML) estimates of movMF model developed in~\cite{banerjee2005clustering}.
	\subsection{EM-based Estimation of Model Parameters}
	In this section, we summarize the approach for approximating the modal parameters using EM algorithm. We estimate the movMF model iteratively using normalized i-Vectors as detailed in Algorithm~\ref{hard_movmf_algo}. The PDF of a $d$-variate vMF distribution is given by
	%
	\begin{equation}
		f(\bm{x}| \bm{\mu}, \kappa) = c_{d}(\kappa) e^{\kappa \bm{\mu}^{T}\bm{x}}
		\label{eq1}
	\end{equation}
	where $||\bm{\mu}||=1$, $\kappa \ge 0$ and $d \ge 2$. Here, input data lies on unit hypersphere $\bm{x} \in \mathbb{S}^{d-1}$ and $(\cdot)^{T}$ denote transpose operation. Normalizing constant, $c_{d}(\kappa)$ is expressed as
	%
	\begin{equation}
		c_{d}(\kappa) = \frac{\kappa ^{\frac{d}{2} -1}}{(2\pi)^{\frac{d}{2}} I_{\frac{d}{2}-1}(\kappa)}
		\label{eq2}
	\end{equation}
	where $I_{r}(\cdot)$ is modified Bessel function of first-kind and order $r$. PDF $f(\bm{x}| \bm{\mu}, \kappa)$ has two parameters, mean direction-vector $\bm{\mu}$, and concentration parameter $\kappa$. The $\kappa$ indicate how strongly the normalized i-Vectors drawn according to $f(\bm{x}| \bm{\mu}, \kappa)$ distribution lie near the mean direction-vector on unit hypersphere. Large $\kappa$ shows substantial concentration near $\bm{\mu}$. 
	\begin{figure*}[!t]
		\centering
		\includegraphics[width=460bp]{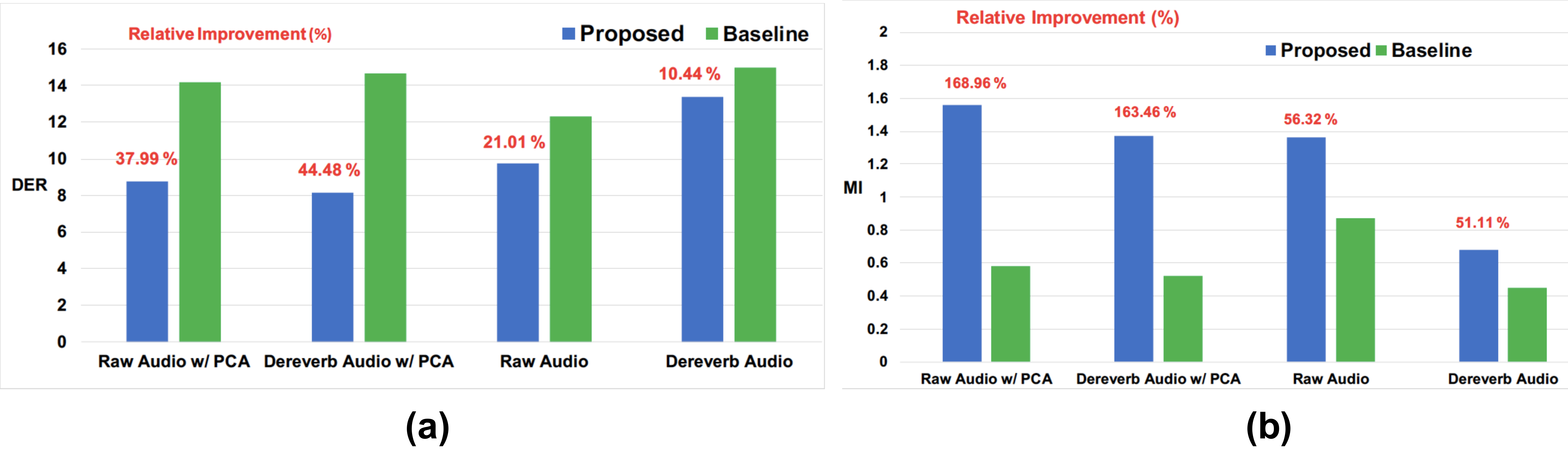}
		\caption{PLTL results: (a) Diarization error rate (DER) for proposed and baseline approaches. We used raw (original) and dereverbed audio in our experiments. The "w/ PCA" denotes PCA-based dimension reduction of i-Vectors to 51 dimensions before length-normalization. Relative reduction (\%) in DER with respect to baseline is shown in red color on top of each bar. (b) Frame-wise mutual information (MI) for proposed and baseline approaches. Relative increase (\%) in MI with respect to baseline is shown in red color above each bar. }
		\label{fig_PLTL}
	\end{figure*}
	%
	Consider a mixture of $N_{c}$ $d$-variate vMF distributions as a generative model for normalized i-Vectors from an audio recording. Let $f_{h}(\bm{x} | \bm{\theta_{h}})$ denote the $h$-th vMF distribution in mixture model and its parameter-vector is $ \bm{\theta_{h}} = (\bm{\mu_{h}}, {\kappa_{h}})$ for $1 \le h \le N_{c}$. Then, the PDF of this mixture model is given as
	%
	\begin{equation}
		f(\bm{x}| \bm{\Theta}) = \sum_{h=1}^{N_{c}}\alpha_{h} f_{h}(\bm{x}|\bm{\theta_{h}})
		\label{eq11}
	\end{equation}
	%
	where $\bm{\Theta} = \{ \alpha_{1}, \cdots, \alpha_{N_{c}}, \bm{\theta_{1}}, \cdots,\bm{\theta_{N_{c}}}\}$; and $\alpha_{h}$ are non-negative weights such that $\sum_{h=1}^{N_{c}} \alpha_{h}=1$. Let $\bm{\chi}=\{\bm{\chi_{1}}, \cdots, \bm{\chi_{n}}\}$ be the stream of normalized i-Vectors modeled with mixture-model in Eq.~\ref{eq11}. Let $\bm{\zeta} = \{ \zeta_{1},  \cdots, \zeta_{n}\}$ be the corresponding set of hidden variables that indicate the component-vMF distribution from which an i-Vector is sampled. Particularly, $\zeta_{i} =h$ if $\bm{\chi_{i}}$ is sampled from distribution $f_{h}(\bm{x} | \bm{\theta_{h}})$. In terms of hidden-variable vector $\bm{\zeta}$, the log-likelihood (LL) of $n$ observed i-Vectors is given by
	\begin{equation}
		\text{ln} \{ P ( \bm{\chi}, \bm{\zeta} | \bm{\Theta}) \} = \sum_{i=1}^{n} \text{ln} \{ \alpha_{\zeta_{i}} f_{\zeta_{i}}(\bm{\chi}_{i} | \bm{\theta}_{\zeta_{i}})\}.
		\label{eq12}
	\end{equation}
	%
	For a given $(\bm{\chi}, \bm{\Theta})$, it is possible to estimate the most likely conditional distribution of $\bm{\zeta} |( \bm{\chi} , \bm{\Theta})$, and this forms the Expectation-step in EM algorithm. Next, we use the EM algorithm for maximizing the expectation of Eq.~\ref{eq12} with constraints $\bm{\mu_{h}}^{T}\bm{\mu_{h}} =1$ and ${\kappa_{h}} \ge 0$. As a result, we get the following expressions for movMF model parameters~\cite{banerjee2005clustering}:
	\begin{equation}
		\alpha_{h} = \frac{1}{n} \sum_{i=1}^{n} p(h|\bm{\chi_{i}}, \bm{\Theta}), 
		\label{eq13}
	\end{equation}
	\begin{equation}
		\bm{r_{h}} = \sum_{i=1}^{n} \bm{\chi_{i}} \  p(h|\bm{\chi_{i}}, \bm{\Theta}), 
		\label{eq14}
	\end{equation}
	\begin{equation}
		\hat{\bm{\mu}}_{h} = \frac{\bm{r_{h}}}{||\bm{r_{h}} ||} , 
		\label{eq15}
	\end{equation}
	\begin{equation}
		\frac{I_{\frac{d}{2}}(\hat{\kappa}_{h})}{I_{\frac{d}{2}-1}(\hat{\kappa}_{h})} = \frac{||\bm{r_{h}} ||}{\sum_{i=1}^{n} p(h| \bm{\chi_{i}}, \bm{\Theta})}.
		\label{eq16}
	\end{equation}
	The Eqs.~\ref{eq15} and~\ref{eq16} correspond to Maximization-step in EM algorithm providing expressions for modal parameters. Next, given these parameters we consider updating the distribution of $\bm{\zeta} |( \bm{\chi} , \bm{\Theta})$ (i.e., an Expectation-step) to maximize the likelihood of i-Vectors given the estimated parameters from Eqs.~\ref{eq15} and~\ref{eq16}.
	
	%
	%
	%
	Since computing $\hat{\kappa}$ involves ratio of Bessel functions (see Eq.~\ref{eq16}), it is not possible to obtain an analytic solution. Various numerical and/or asymptotic methods are used for approximating $\hat{\kappa}$. In this paper, we used the $\hat{\kappa}$ estimates developed in~\cite{banerjee2005clustering}. It is given as 
	%
	%
	%
	%
	%
	%
	%
	%
	\begin{equation}
		\hat{\kappa} = \frac{\bar{r} d -\bar{r}^3}{1 - \bar{r}^2}.
		\label{eq10}
	\end{equation}
	%
	%
	%
	%
	%
	\begin{figure}[!t]
		\centering
		\includegraphics[width=230bp, trim={10mm 15mm 25mm 5mm},clip]{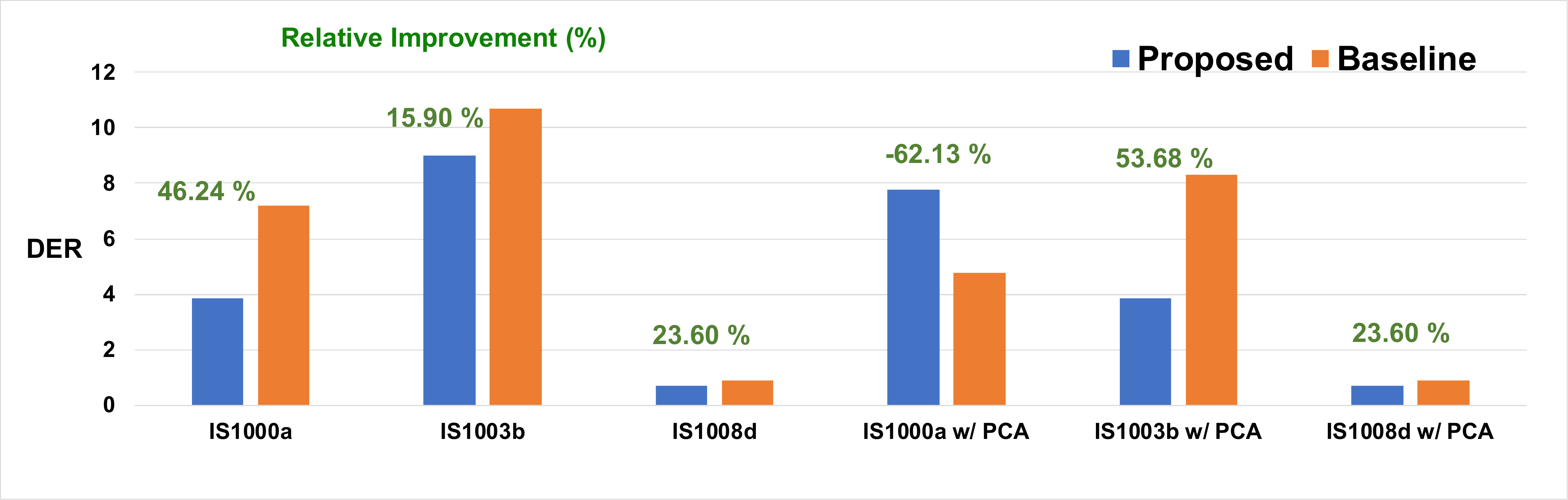}
		\caption{AMI (three-meetings subset) results: DER for proposed and baseline approaches. The "w/ PCA" case refers to PCA-based dimension reduction of i-Vectors to 65 before length-normalization. Relative reduction (\%) in DER with respect to baseline is shown in green color above each bar.}
		\label{fig_AMI}
	\end{figure}
	%
	\subsection{Proposed Speaker Clustering: Algorithm~\ref{hard_movmf_algo}}
	%
	We outline the proposed approach in Algorithm~\ref{hard_movmf_algo}. It outputs the maximum-likelihood (ML) estimates of movMF parameters. This method essentially iterate over two steps in standard EM algorithm until it converges. In each Expectation-step, i-Vectors are hard-assigned to a cluster. The distribution of hidden variables is given by~\textit{Step 9} of Algorithm~\ref{hard_movmf_algo}. The i-Vectors are hard-assigned to a unique cluster on the basis of derived posterior distribution. Cluster assignment is done by computing the $\arg\max$ over posteriors for each i-Vector (\textit{Step 9}). 
	
	Next in Maximization-step of EM, the model parameters of component-vMFs are updated using the component-vMF distributions given the i-Vectors (\textit{Step 12} to \textit{Step 19} in Algorithm~\ref{hard_movmf_algo}). In the proposed hard-assignment approach, the posterior probabilities are restricted to have only binary i.e., 0 or 1 values. With hard-assignments, the distribution of hidden-variables is restricted to assume probability value 1 for some mixture component and 0 for all others. This strategy maximizes a lower-bound on incomplete log-likelihood (LL) of i-Vectors~\cite{banerjee2005clustering}. In other words, expectation over distribution $q(\cdot)$ lower bounds the LL of i-Vectors. 
	
	Upon convergence, the Algorithm~\ref{hard_movmf_algo} outputs the movMF model parameters and $N_{c}$ clusters of i-Vector data (\textit{Step 20}). The hard-assignments in~\textit{Step 9} reduce the computational complexity as posterior probabilities are binary values. Proposed clustering approach in Algorithm~\ref{hard_movmf_algo} requires only $\mathcal{O}(N_{c})$ computations in each EM iteration. It need to store only the cluster assignments for all i-Vectors (i.e., $n$ integers). These two facts make the proposed speaker clustering computationally efficient and scalable. 
	%
	%
	%
	%
	\section{Experiments, Results~\& Discussions}
	%
	\subsection{Baseline System}
	%
	The cosine similarity was previously used for comparing i-Vectors in K-means and mean-shift clustering~\cite{castaldo2008stream,senoussaoui2014study}. In this paper, we adopt spherical K-means (with cosine similarity) as baseline approach~\cite{zhong2005efficient}. Spherical K-means projects the estimated cluster centroids onto unit hypersphere at the end of each maximization-step unlike conventional K-means. Spherical K-means is a special case of proposed movMF-based speaker clustering. If we impose all mixture weights ($\alpha_{h} \text{ for } 1 \le h \le N_{c} $) to be equal and all concentration parameters ($\kappa_{h} \text{ for } 1 \le h \le N_{c} $) to be equal (with any value), then the proposed movMF-based speaker clustering becomes equivalent to spherical K-means. 
	\subsection{Evaluation Data}
	%
	\subsubsection{AMI Corpus}
	%
	Augmented Multi-party Interaction (AMI) corpus has multi-modal data from meeting scenarios with reference speaker annotations. We derived our AMI evaluation set from audio recordings of multi-speaker interactions in three meetings. We used~\textit{headset audio} data for our experiments. We chose a three-meetings subset containing sessions~\verb!IS1000a! (26 min.), \verb!IS1003b! (27 min.) and~\verb!IS1008d! (24 min.) from the popular 12-meetings subset of AMI corpus~\cite{mccowan2005ami, gonina2011fast}. Each of the three meetings has four speakers discussing the design of a new remote control device. 
	%
	%
	%
	%
	\subsubsection{CRSS-PLTL Corpus}
	\label{sec:pltl_data}
	%
	Each PLTL session has a peer-leader to facilitate discussions among 6-8 students. PLTL model is known to enhance student's learning~\cite{cracolice2001peer}. Typically, peer-leaders have previously passed the course with good grades. It is implemented for undergraduate STEM courses in U.S. universities. In association with \textit{UT-Dallas Student Success Center}, we collected the CRSS-PLTL corpus~\cite{dubey2017csl,dubey2016interspeech,dubey2016slt,dubey2018tasl,dubey2018pltlis} from five teams attending a chemistry course over eleven weeks. The weekly sessions were organized for approximately 70-80 minutes. Each participant wore a LENA device (with not-so-close microphone) for collecting naturalistic audio~\cite{dubey2018pltlis,hansen2018ldnn}. In this manner, we collected multi-stream audio for each session (number of streams was same as total participants). The salient features of this data are: (i) many segments with overlapped-speech; (ii) short conversational-turns; (iii) multiple noise-sources; and (iv) significant reverberation. These factors made PLTL speaker diarization challenging. In this paper, we choose the channel corresponding to PLTL leader for diarization evaluation. 
	%
	\subsection{Evaluation Metric}
	%
	We used two evaluation metrics namely: (i) diarization error rate (DER); and (ii) frame-level mutual information (MI) for scoring the system-output with respect to reference annotation. DER was introduced in the NIST Rich Transcription Spring 2003 evaluation (RT-03S). It is defined as the total percentage of reference speaker-time that is not correctly attributed to a speaker. Mathematically, DER~\cite{nistder} is given as:
	%
	\begin{equation}
		\text{DER} = \frac{\phi_{fa} + \phi_{miss} + \phi_{err} }{\phi_{total}},
		\label{eqn_der}
	\end{equation}
	%
	where $\phi_{total}$ is the total time of all reference-segments, $\phi_{fa}$ is the system speaker-time not attributed to the reference speaker, $\phi_{miss}$ is the total reference speaker-time not attributed to a system speaker, and $\phi_{err}$ is the total reference speaker-time attributed to a wrong speaker. Unlike NIST RT evaluations~\cite{nistrt09}, we do not apply any forgiveness collar to the reference human annotations prior to scoring. We consider overlapped-speech as an additional cluster while scoring the system outputs and during human annotations. We adopted the NIST md-eval scoring script (version-22) for DER computations~\cite{nistder}.
	%
	%
	
	MI quantifies the statistical similarity between frame-level system output and ground-truth. We incorporate overlapped-speech in MI computation. First of all, both ground-truth and system output are converted to 10ms frame-level labels. Then, the frame-level MI (in bits) between system output and ground-truth is mathematically defined as:
	%
	\begin{equation}
		\text{MI} = \sum_{i=1}^{R} \sum_{j=1}^{S}  \frac{n_{ij}}{N} \log_{2} \frac{n_{ij} N}{r_{i} s_{j}}, 
	\end{equation}
	where $R$, $S$ are the number of reference and system clusters, respectively; $n_{ij}$ is the number of frames assigned to $i$-th reference and $j$-th system cluster; $r_{i}$, $s_{j}$ are the number of frames assigned to $i$-th reference, and $j$-th system cluster, respectively; and $N$ is the total number of frames. We compute MI values using the scoring scripts from the First DIHARD Challenge Evaluation~\cite{dihard_scoring}. 
	%
	\subsection{Results}
	%
	In this study, the PLTL data contains audio of the peer-leader's channel from a 80-minute session with seven students. We obtain the ground-truth segmentation and speaker labels from human annotators. Figure~\ref{fig_PLTL} compares the performance of proposed approach with baseline on PLTL data. Sub-figure~\ref{fig_PLTL} (a) illustrates the DER while (b) shows the frame-level MI for raw and dereverbed audio. The majority of PLTL speaker-turns were less than one-second (though few lasted over two-seconds), we chose 75 (lower) dimensional i-Vectors. We repeated all experiments with PCA for reducing the i-Vector dimension to 51. This PLTL audio contains significantly 10\% overlapped-speech that was incorporated as a separate cluster during evaluation. Thus, the number of cluster, $N_{c}$= 9 that includes peer-leader, seven students and overlapped-speech. The proposed approach has improved performance in terms of lower DER and higher MI values compared to baseline. The consistent improvement in all cases with original and enhanced audio, with or without PCA validate the efficacy of movMF model for robust speaker clustering. We have similar observation on three-meetings subset of AMI data as illustrated in Fig.~\ref{fig_AMI}. We included only DER for AMI data to avoid presenting too many results. The CRSS-PLTL audio has higher levels and more varied forms of distortions as compared to AMI corpus resulting in a challenging diarization scenario. The proposed clustering approach has the ability to adapt the concentration parameter $\kappa$ for each component-vMF distribution in the mixture-model. This created a flexible modeling of normalized i-Vectors that is substantially better than spherical K-means as spherical K-means do not estimate the weight or concentration parameters unlike movMF model. The movMF clustering do a better job by taking advantage of the concentration estimates for each component-vMF distribution. 
	\section{Conclusions}
	This paper proposed modeling of length-normalized i-Vectors with a mixture of multi-variate von Mises-Fisher distributions (movMF). Standard EM algorithm was used for estimating the model parameters. The normalized i-Vectors are high-dimensional data lying on unit hypersphere that facilitated reasonable approximation of movMF model parameters. In general, movMF parameters do not have closed form solutions. The model parameters are leveraged for robust speaker clustering. The evaluation data was derived from naturalistic CRSS-PLTL and AMI meeting corpus. Accurate modeling resulted in improved performance of movMF speaker clustering compared to baseline spherical K-means. 
	%
	%
	\bibliographystyle{IEEEtran}
\bibliography{mvf_CR}
	%
	%
\end{document}